# EMT-NET: EFFICIENT MULTITASK NETWORK FOR COMPUTER-AIDED DIAGNOSIS OF BREAST CANCER

*Jiaqiao Shi[1], Aleksandar Vakanski[1], Min Xian[1*], Jianrui Ding[2], Chunping Ning[3]*

[1]Department of Computer Science, University of Idaho, Idaho Falls, ID 83401, USA
[2]School of Computer Science and Technology, Harbin Institute of Technology, Weihai, China
[3]Department of Ultrasound, Medical College at Qingdao University, Qingdao, China

## ABSTRACT

Deep learning-based computer-aided diagnosis has achieved unprecedented performance in breast cancer detection. However, most approaches are computationally intensive, which impedes their broader dissemination in real-world applications. In this work, we propose an efficient and light-weighted multitask learning architecture to classify and segment breast tumors simultaneously. We incorporate a segmentation task into a tumor classification network, which makes the backbone network learn representations focused on tumor regions. Moreover, we propose a new numerically stable loss function that easily controls the balance between the sensitivity and specificity of cancer detection. The proposed approach is evaluated using a breast ultrasound dataset with 1511 images. The accuracy, sensitivity, and specificity of tumor classification is 88.6%, 94.1%, and 85.3%, respectively. We validate the model using a virtual mobile device, and the average inference time is 0.35 seconds per image.

## 1. INTRODUCTION

Breast cancer is one of the most common cancers and leading causes of death among women worldwide. Alone in 2019 in the U.S., around 268,600 new cases of invasive breast cancer and 62,930 new cases of non-invasive breast cancer in women were expected to be diagnosed [1]. Many CAD systems have been clinically tested and demonstrated their ability to improve the diagnostic sensitivity, specificity, and efficiency of breast cancer diagnosis [2].

Prominent progress in CADs has been achieved by using deep learning techniques in recent years. Applying deep learning with Convolutional Neural Network (CNN) to classify breast ultrasound image has gained popularity. To date, many studies have reported the usefulness of deep learning for diagnostic imaging of breast masses with ultrasound. Shi et al. [3] utilized the stacked deep polynomial network to learn textural features from 100 malignant and 100 benign masses on ultrasound images. Stoffel et al. [4] focused on phyllodes tumor and fibroadenoma classification using deep learning and obtained great negative predictive value (NPV: 100%) and promising accuracy (AUC: 0.73). Tumor samples of 4254 benign and 3154 malignant cases were employed to train the deep CNN in [5], where GoogLeNet architecture was used; and it reported an accuracy of 91%, a sensitivity of 86%, a specificity of 93%, and an AUC over 0.90. Shin et al. [6] achieved improved classification performance through a pretraining CNN using the ImageNet and fine-tuning the network on a BUS dataset.

BUS image segmentation is an essential step in a CAD system, which aims to extract tumor regions from normal breast tissues. Due to the speckle noise, poor image quality, and variable tumor shapes and sizes, achieving accurate BUS image segmentation is challenging [7]. There still are common and fundamental issues in BUS segmentation, e.g., the denoising and preserving edges, operator-dependency, reproducibility, and domain knowledge incorporation. CNN-based models, such as Fully Convolutional Network (FCN) [8] and U-Net [9], have widely applied in BUS image segmentation. An RDAU-NET [10] improved the regular U-Net and achieved a precision of 88.6%, sensitivity of 83.2%, and F1 score of 0.85. By combining a dilated fully convolutional network with a phase-based active contour model for automatic tumor segmentation, the constructed model is reported to exhibit high robustness, accuracy, and efficiency [11]. Kumar et al. [12] introduced their multi-U-net algorithm for automatic and effective breast masse segmentation. It achieved a mean Dice coefficient of 82%, a true positive rate of 84%, and a false positive rate of 1%. Shareef et al. [13] proposed a novel architecture called Small Tumor-Aware Network (STAN), where both rich context information and high-resolution image features are integrated and is proved to improve segmentation performance on tumors with different size. ESTAN is then proposed in [20], where two encoders are introduced to extract and fuse image context information at various scales, and row-column-wise kernels are utilized in the encoder to adapt to the breast anatomy. Vakanski et al. [21] proposed a salient attention approach that introduces attention blocks into a U-Net architecture and learns feature representations that prioritize spatial regions with high saliency levels.

* Correspondence to Min Xian (mxian@uidaho.edu). Research reported in this publication was supported, in part, by the National Institute of General Medical Sciences of the National Institutes of Health under Award Number P20GM104420. The content is solely the responsibility of the authors and does not necessarily represent the official views of the National Institutes of Health.

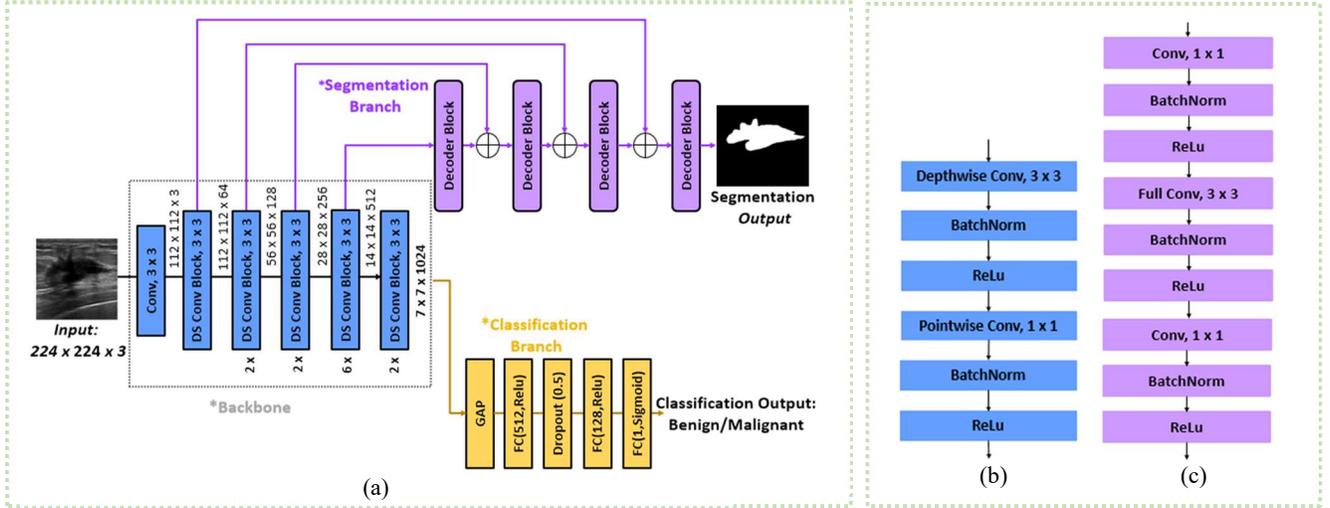

**Fig. 1**. The proposed efficient multitask network. (a) Overall architecture; (b) the depthwise separable convolutional block (DS Conv Block) [14]; and (c) the decoder block [15].

The above discussed deep learning approaches have achieved popularity in breast cancer detection. However, two major challenges exist: (1) The large number of parameters and high computation intensity make it difficult to deploy these approaches to devices with low computing resources; and (2) existing approaches treat the false positive and false negative detections equally, and have no control of the balance of the sensitivity and specificity.

In this work, we overcome the challenges by proposing an efficient multitask network that achieves competitive performance with fewer parameters. It applies an efficient backbone network, and integrates tumor segmentation task into a major tumor classification network to push the major task to perform tumor classification using relevant tumor features. It obtains the tumor classification and segmentation results simultaneously and can avoid deploying two large independent models on mobile devices. Furthermore, we propose a numerically stable weighted cross-entropy loss, which leads to a better control of the trade-off between sensitivity and specificity.

## 2. PROPOSED APPROACH

### 2.1. Efficient multitask network (EMT-Net) architecture for breast cancer diagnosis

Training two separate tumor segmentation and classification models using the same dataset could achieve breast tumor detection and classification sequentially, but it is inefficient from the standpoint of computation cost and model storage cost. In real clinic practice, the segmented lesion help determine tumor categories, assess the degree of malignancy, and plan follow-up treatments. A multitask (MT) learning architecture with both tumor classification and segmentation tasks is proposed in this section, and the classification is the primary task, and segmentation is the auxiliary task. The motivation behind this design is to leverage useful information from related secondary task to improve the generation of the primary task. Fig. 1 shows both the overall network architecture with classification and segmentation branches, along with the detail for layers in the encoder linking with corresponding layers in the decoder.

**Backbone network**. The pretrained MobileNet (V1) [14] is utilized as the shared backbone network (encoder) to efficiently extract the local feature representations from the input ultrasound images. Our choice of MobileNet as the backbone network is that its separable convolutions substantially improve computation efficiency. The detail component of each depthwise separable convolution (DS Conv Block) is shown in Fig. 1(b). In the proposed network, only the pretrained convolutional layers are employed, while its original output full connected (FC) layers are not included. The output feature maps have a shape of $7 \times 7 \times 2024$.

**Tumor classification** is the primary task in the proposed MT network, and it classifies tumors into benign or malignant categories. It accepts input from the backbone network, which has a shape of $7 \times 7 \times 2024$, followed by a Global Average Pooling (GAP) layer, then two fully connected (FC) layers using ReLu activation and with 512 and 128 nodes, respectively. A dropout layer with a dropout rate of 0.5 is inserted after the first new dense layer. The last one-node FC layer with the Sigmoid activation defines the final output layer. The output of the tumor classification branch is either benign (0) or malignant (1) tumor.

**Tumor segmentation.** The segmentation branch takes advantages of skip connections, and achieves efficient dense prediction using four decoder blocks. The decoder details are shown in Fig. 1 (c). The decoder block follows the design in [15] is applied in the segmentation branch to achieve pixel-wise semantic segmentation has a 3 × 3 full-convolution layer inserted between two 1 × 1 convolution layers. Batch normalization followed by ReLU activation is applied to the output of each convolution layer. As shown in Fig. 1(a), the output of the first three convolutional layers are directly combined with the output of the corresponding decoder layers with the same feature map shape. There are three skip combinations between the encoder and segmentation branch with the shape of 112 × 112 × 64, 56 × 56 × 128, and 28 × 28 × 256, sequentially. The combination operation is the element-wise addition. Through these three skip connections, spatial information could be recovered with fewer parameters, and thus the entire network keeps efficient.

## 2.2. Numerically-stable weighted binary cross-entropy

In CAD systems for disease diagnose, sensitivity is considered to be more important than other performance metrics. Sensitivity measures a model's ability to find disease cases out of normal cases. Poor sensitivity of a model will result in the missing classification of many positive cases to negative cases (high false negative). In actual clinic practice of breast cancer diagnosis, the false negatives will lead to the delayed detection of breast cancer and extremely lower the survival chance of patients. To make a better trade-off between the sensitivity and specificity while not decreasing the overall accuracy of the model, we propose the numerical-stable weighted binary cross-entropy (WBCE) as the loss function for the classification branch. The standard WBCE is defined by

$$L_{WBCE} = -\frac{1}{M}\sum_{i=1}^{M}(W_p \cdot Y_i \cdot \log(H_i) + (1 - Y_i) \cdot \log(1 - H_i)) \quad (1)$$

where M denotes the number of training examples; $Y_i$ is the target label of the $i$th training sample; $H_i$ is the prediction; and the weight $W_p$ is added as a multiplicative coefficient for the positive labels term. The additional weight $W_p$ with value large than 1 will increase the loss for false negatives; hence minimizing the loss function will increase the sensitivity for cancer detection. To ensure stability and avoid overflow, we implement the numerical-stable WBCE loss function ($L_{NS-WB}$) which is defined by

$$L_{NS-WB} = \frac{1}{M}\sum_{i=1}^{M}((1 - Y_i) \cdot Z + K \cdot \log(1 + e^{-Z})) \quad (2)$$

$$Z = \log\left(\frac{H_i}{1 - H_i}\right) \quad (3)$$

$$K = 1 + Y_i \cdot (W_p - 1) \quad (4)$$

where $Z$ and $K$ are two intermediary elements; $Z$ denotes the logits converted back from the predicted probability $H_i$, and $K$ is converted from the positive term coefficient.

## 2.3. Final multitask loss

In our experiments, we implement the Dice Loss ($L_{Dice}$) (Eq. 5) for the segmentation branch.

$$L_{Dice} = 1 - \frac{2 \cdot \sum_{i=1}^{M}(Y_i \cdot H_i)}{\sum_{i=1}^{M} Y_i + \sum_{i=1}^{M} H_i} \quad (5)$$

Among the two tasks, classification is our primary task, and we use the parameter $W_{clf}$ to weight the contribution of the classification loss. When the $W_{clf}$ is set to be values greater than 1, the network training will focus more on the performance improvement of classification. The final loss function of the proposed network has two controlled weight parameters. One is $W_p$ as the positive term coefficient inside the $L_{NS-WBCE}$, and the other one is $W_{clf}$ as the classification loss coefficient multiplying with the classification loss. The final loss ($L_{DW-M}$) is the weighted sum of $L_{NS-WBCE}$ and $L_{Dice}$.

$$L_{DW-M} = W_{clf} \cdot L_{NS-WBCE} + L_{Dice}. \quad (6)$$

## 3. EXPERIMENTAL RESULTS

### 3.1. Dataset and evaluation metrics

We use breast ultrasound images from four public datasets, the BUSIS dataset [16], the Dataset B [17], the Thailand dataset [18] and the BUSI dataset [19].

All the introduced classification and segmentation models require input with fixed and equal width and height, while all the images in the datasets are originally rectangle shaped. Padding strategy is introduced as a data preprocessing to transform original images from rectangle to square shape without having tumor distortion.

Metrics for evaluating the classification include accuracy (ACC), sensitivity (SEN), and specificity (SPE). Metrics for evaluating the segmentation include the Dice's coefficient (DSC) and Intersection over Union (IoU).

### 3.2. The effectiveness of the weighted cross-entropy

The effectiveness of the weight $W_p$ added for positive labels term in $L_{NS-WBCE}$ is measured on a list of 14 different values from 1 t o 5 with step size of 0.5, and 6 to 10 with step size of 1. Controlled experiments are conducted on single-task classification (Single-CLF) scenario, where the Single-CLF model is the backbone network followed with the tumor classification branch. The K-fold cross-validation is implemented with random seed value 42 to shuffle and split data. With K being 4, the dataset is split into training set of 75%, validation set of 12.5% and test set of 12.5%. Fig. 2. plots the average ACC, SEN and SPE calculated over the 4 folds of test set, for each positive coefficient value. Starting from the standard BCE loss where the positive weight is 1, the SEN gets improved along with the weight value increases.

The trend of ACC/SEN/SPE curves in Fig. 2 prove that, higher weight values for the positive label term strongly penalizes the false negatives and compensates the class

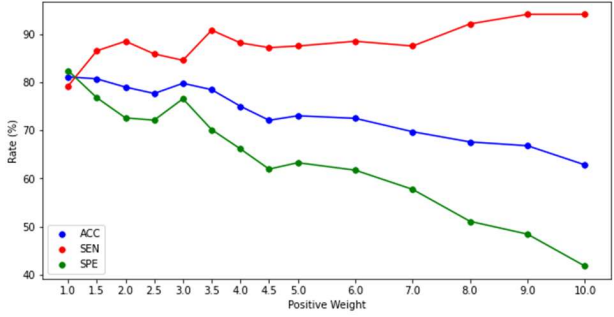

**Fig. 2.** K-fold average ACC, SEN and SPE with different $W_p$.

**Table 1.** Classification and Segmentation performance on test set of the proposed network, Single-CLF and Single-SGM.

| Models | EMT-Net | Single-CLF | Single-SGM |
|---|---|---|---|
| **Size** | **60**MB | 30MB | 54MB |
| **Parameters** | **5.1**M | 3.8M | 4.5M |
| ACC (%) | **88.6** | 81.1 | - |
| SEN (%) | **94.1** | 79.2 | - |
| SPE (%) | **85.3** | 82.4 | - |
| DSC | 0.81 | - | **0.84** |
| IoU | 0.73 | - | **0.76** |

* The best score in each row is highlighted using bold font.

imbalance problem. Since SPE starts having a sharp fall when $W_p$ is greater than 3, thus overall ACC gets decreased largely.

### 3.3. The effectiveness of the EMT-Net

Controlled experiments are performed to evaluate the classification and segmentation performance of the constructed EMT-Net. The entire dataset is split into a training set of 70%, a validation set of 15%, and a test set 15% with random seed 42.

The baseline model is the proposed EMT-Net with $W_{clf}$ as 1.5 and $W_p$ as 3. We compare its performance with a single-task classification (Single-CLF) and segmentation model (Single-SGM), constructed by MobileNetV1 with the classification and segmentation branch, respectively. As shown in Table 1, the proposed EMT-Net dramatically boosts the classification performance with a slight drop in segmentation performance. In addition, the number of parameters is only 61.5% of the parameters of the combined Single-CLF and Single-SGM models. The model size is only 71.4% of the combined models.

### 3.4. Weights tuning

The total empirical loss is minimized with the trade-off of learning individual tasks. In addition, to improve the SEN as much as possible while keeping ACC and SPE relatively high, an appropriate value for $W_p$ needs to be explored. Controlled experiments on combinations of $W_{clf}$ and $W_p$ values, both from 1 to 3 with step size 0.5, are conducted.

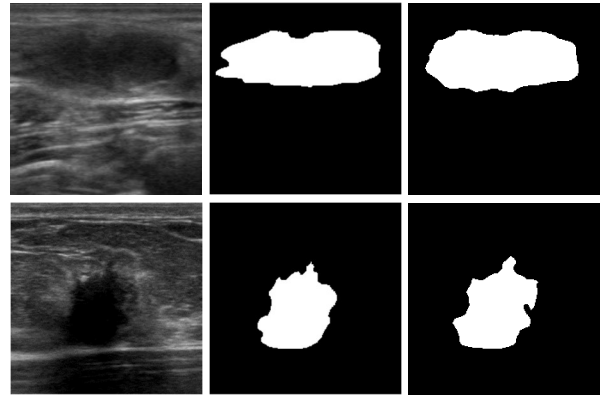

**Fig. 3.** Examples of two original images (left), ground truths (middle), and segmentation results (right) of the proposed approach.

There are total 25 combinations of $W_{clf}$ and $W_p$. With $W_p$ being 3 and $W_{clf}$ being 1.5, we achieve the best trade-off between SEN and SPE. Fig. 3 shows the segmentation results of two breast ultrasound images using the proposed network.

### 3.5. Test using a virtual mobile device

We develop a mobile application using Android Studio 3.2 to test the effectiveness of our multitask network on mobile devices. Android SDK with version 23 and an Android Emulator (simulate a physical Android mobile device on the computer) are used. A Google Pixel 2 API 28 with CPU x86 and 9.2 GB size on disk is downloaded as the instance of the Android virtual device. The final trained model is converted into TensorFlow Lite format. The average inference time to simultaneously segment and classify a breast ultrasound image is only about 0.35 seconds.

### 4. CONCLUSION

In this paper, we propose the EMT-Net for efficient breast cancer diagnosis on mobile devices. The network has a model size of 20MB after converting to Tensorflow Lite format, and could support fast breast cancer detection using mobile devices. The proposed network integrates one primary task (tumor classification) and one auxiliary task (tumor segmentation), and significantly outperforms the tumor classification performance of a single-task network. The newly proposed numerically stable weighted cross-entropy loss is a flexible tool to balance the sensitivity and specificity for breast cancer detection. We can easily increase the model's sensitivity by increasing the positive weight ($W_p$). In future work, we will investigate the viability of the proposed multitask network on a variety of breast cancer imaging modalities, and explore the further improvement of the sensitivity and specificity of the proposed model.